\documentclass[12pt]{article}
\usepackage{amsmath,amsthm}
\usepackage{eufrak}

\newcommand{\ar}{\renewcommand{\arraystretch}{1}} 
\DeclareMathAlphabet{\bb}{U}{msb}{m}{n}
\gdef\C{\bb C}

\gdef\R{\bb R}

\DeclareMathOperator{\spin}{{\bf Spin}}

\newcommand{\re}{\mbox{\rm Re}\,}
\newcommand{\im}{\mbox{\rm Im}\,}

\newcommand{\cP}{{\cal P}}
\newcommand{\cL}{\mathcal{L}}
\newcommand{\cM}{{\cal M}}

\newcommand{\sA}{{\sf A}}
\newcommand{\sB}{{\sf B}}

\newcommand{\sJ}{{\sf J}}

\newcommand{\sX}{{\sf X}}
\newcommand{\sY}{{\sf Y}}
\newcommand{\sK}{{\sf K}}

\newcommand{\bk}{{\bf k}}
\newcommand{\bp}{{\bf p}}
\newcommand{\bx}{{\bf x}}
\newcommand{\by}{{\bf y}}
\newcommand{\bz}{{\bf z}}
\newcommand{\bB}{{\bf B}}

\newcommand{\bE}{{\bf E}}

\newcommand{\fM}{\mathfrak{M}}

\newcommand{\fL}{\mathfrak{L}}
\newcommand{\fT}{\mathfrak{M}}

\newcommand{\fg}{\mathfrak{g}}

\newcommand{\balpha}{\boldsymbol{\alpha}}
\newcommand{\cl}{C\kern -0.2em \ell}

\newcommand{\hypergeom}[5]{\mbox{$
_#1 F_#2\left.
\!\!
\left(
\!\!\!\!
\begin{array}{c}
\multicolumn{1}{c}{\begin{array}{c}
#3
\end{array}}\\[1mm]
\multicolumn{1}{c}{\begin{array}{c}
#4
\end{array}}\end{array}
\!\!\!\!
\right|\displaystyle{#5}\right)
$}
}

\newcommand{\lf}{\left\{}
\newcommand{\rf}{\right\}}
\newcommand{\tg}{\tan}
\newcommand{\ch}{\cosh}

\newcommand{\tnh}{\tanh}

\newcommand{\ctg}{\cot}
\begin{document}


%
%

\title{MAXWELL FIELD ON THE POINCAR\'{E} GROUP}

\author{V. V. Varlamov\\
{\small\it Department of Mathematics, Siberia State University of Industry,} \\
{\small\it Kirova 42, Novokuznetsk, 654007,
Russia}}


\date{}
\maketitle


\begin{abstract}
The massless field of spin 1 is defined on the eight-dimensional
configuration space; this space is a direct product of Minkowski space
and of a two-dimensional complex sphere. Field equations for the spin-one
field are derived from a Dirac-like Lagrangian separately for the
translation group and Lorentz group parts. It is shown that a Dirac form
of Maxwell equations (so-called Majorana-Oppenheimer formulation of
electrodynamics) follows directly from the field equations of translation
group part. The photon field is realized via Biedenharn type functions
on the Poincar\'{e} group. This allows us to consider both Dirac and
Maxwell fields on an equal footing, as the functions on the
Poincar\'{e} group.

\end{abstract}

\section{Introduction}
Historically, an analysis of the Maxwell equations with respect to Lorentz
transformations is one of the first objects of relativity theory. As known, any
quantity which transforms linearly under Lorentz transformations is a spinor.
For that reason spinor quantities are considered as fundamental in quantum
field theory and basic equations for such quantities should be written
in a spinor form. The spinor form of Dirac equations was first given by
Van der Waerden \cite{Wa29}, he showed that Dirac's theory can be expressed
completely in this form (see also Ref. 2). In turn, a spinor
formulation of Maxwell equations was studied by Laporte and 
Uhlenbeck \cite{LU31}. In 1936, Rumer \cite{Rum36} showed that spinor forms of Dirac and
Maxwell equations look very similar\footnote{Of course, there is no an
equivalence between Dirac and Maxwell equations as it claimed recently by
Campolattaro {\em et al.} \cite{Cam90,VR93} (see also a discussion
concerning this question \cite{Gsp02}). A principal difference between
these equations lies in the fact that Dirac and Maxwell fields have
different spintensor dimensionalities. These fields are transformed 
correspondingly within $(1/2,0)\oplus(0,1/2)$ and $(1,0)\oplus(0,1)$
finite dimensional representations of the Lorentz group.}.
Further, Majorana \cite{Maj} and Oppenheimer \cite{Opp31} proposed to
consider the Maxwell theory of electromagnetism as the wave mechanics
of the photon. They introduced a wave function of the form
$\psi=\bE-i\bB$ satisfying the massless 
Dirac-like equations\footnote{In contrast to the 
Gupta-Bleuler method \cite{Gup50,Ble50}, where the nonobservable
four-potential $A_\mu$ is quantized, the main advantage of the
Majorana-Oppenheimer formulation of electrodynamics lies in the fact that it
deals directly with observable quantities, such as the electric and
magnetic fields.}. Maxwell equations in the Dirac form considered during long
time by many
authors \cite{Arc55}--\cite{Ljo88}. 
The interest to the Majorana-Oppenheimer formulation
of electrodynamics has grown in recent years \cite{Sal90}--\cite{Esp98}.

It is widely accepted in modern theoretical physics that the majority
of ``elementary" particles have a very complicated internal structure, that is,
elementary particles seem to be spatially extended. For that reason
realistic particles cannot be considered as point-like objects (of course,
these more realistic fields are nonlocal fields in general). Aside string
models, there is another way for description of spatially extended particles
proposed by Finkelstein \cite{Fin55}, he showed that elementary particle
models with internal degrees of freedom can be described on manifolds larger
then Minkowski space (homogeneous spaces of the Poincar\'{e} group $\cP$).
All the homogeneous spaces of $\cP$, which contain Minkowski space, were
given by Finkelstein \cite{Fin55} and by Bacry and Kihlberg \cite{BK69}.
In 1964, under the influence of Regge pole theory, Lur\c{c}at
suggested to construct quantum field theory on the group manifold of $\cP$
\cite{Lur64}, where one of the main motivations was to give a dynamical
role to the spin. The constructions of quantum fields theories on different
homogeneous spaces of $\cP$ were given in Refs. \cite{Kih70}--\cite{GL01}.

The main goal of the present paper is a synthesis of the two mentioned above
directions (Dirac-like formulation of Maxwell equations and quantum field
theory on the Poincar\'{e} group). The Maxwell field is represented by
Biedenharn type functions \cite{BBTD88} on the group manifold $\cM_{10}$
(this manifold is a direct product of Minkowski space and of the manifold
of the Lorentz subgroup). It is shown that a general form of the
wavefunction inherits its structure from the semidirect product
$SL(2,\C)\odot T_4$ and for that reason the Maxwell field on $\cM_{10}$
is defined by a factorization $\psi(x)\psi(\fg)$, where $x\in T_4$,
$\fg\in SL(2,\C)$. It is obvious that the Dirac-like form of Maxwell
equations should be derived from a Dirac-like Lagrangian.
Using a Lagrangian formalism on the tangent bundle
$T\cM_{10}$ of the manifold $\cM_{10}$, we obtain field equations
separately for the parts $\psi(x)$ and $\psi(\fg)$. Solutions of the field
equations for $\psi(x)$ (Maxwell equations in Dirac-like form) are obtained
via the plane-wave approximation\footnote{As known, exponentials define
unitary representations of the translation subgroup $T_4$. In a sense,
the functions $e^{ikx}$ can be understood as ``matrix elements'' of $T_4$.}.
In turn, solutions of the field equations with $\psi(\fg)$ have been found
in the form of expansions in associated hyperspherical 
functions\footnote{Matrix elements of both spinor and principal series
representations of the Lorentz group are expressed via the hyperspherical
functions \cite{Var034} (see Appendix).}.
\section{Preliminaries}
Let us consider some basic facts concerning the Poincar\'{e} group $\cP$.
First of all, the group $\cP$ has the same number of connected components
as with the Lorentz group. Later on we will consider only the component
$\cP^\uparrow_+$ corresponding the connected component
$L^\uparrow_+$ (so-called special Lorentz group \cite{RF}). As known,
an universal covering $\overline{\cP^\uparrow_+}$ of the group $\cP^\uparrow_+$
is defined by a semidirect product
$\overline{\cP^\uparrow_+}=SL(2,\C)\odot T_4\simeq\spin_+(1,3)\odot T_4$, 
where $T_4$ is a subgroup of four-dimensional translations. 

The each transformation $T_{\balpha}\in\cP^\uparrow_+$ is defined by
a parameter set $\balpha(\alpha_1,\ldots,\alpha_{10})$, which can be 
represented by a point of the space $\cM_{10}$. The space $\cM_{10}$
possesses locally euclidean properties, therefore, it is a manifold called
{\it a group manifold of the Poincare group}. It is easy to see that the
set $\balpha$ can be divided into two subsets,
$\balpha(x_1,x_2,x_3,x_4,\fg_1,\fg_2,\fg_3,\fg_4,\fg_5,\fg_6)$, where
$x_i\in T_4$ are parameters of the translation sugroup,
$\fg_j$ are parameters of the group $SL(2,\C)$. In turn, the transformation
$T_{\fg}$ is defined by a set $\fg(\fg_1,\ldots,\fg_6)$, which can be
represented by a point of a six-dimensional submanifold $\fL_6\subset\cM_{10}$
called {\it a group manifold of the Lorentz group}.

In the present paper we restricted ourselves by a consideration of finite
dimensional representations of the Poincar\'{e} group. The group $T_4$ of
four-dimensional translations is an Abelian group, formed by a direct
product of the four one-dimensional translation groups, the each of which
is isomorphic to additive group of real numbers. Hence it follows that all
irreducible representations of $T_4$ are one-dimensional and expressed
via the exponential. In turn, as it showed by Naimark \cite{Nai58},
spinor representations exhaust all the finite dimensional irreducible
representations of the group $SL(2,\C)$. Any spinor representation of
$SL(2,\C)$ can be defined in the space of symmetric polynomials of the
following form
\begin{gather}
p(z_0,z_1,\bar{z}_0,\bar{z}_1)=\sum_{\substack{(\alpha_1,\ldots,\alpha_k)\\
(\dot{\alpha}_1,\ldots,\dot{\alpha}_r)}}\frac{1}{k!\,r!}
a^{\alpha_1\cdots\alpha_k\dot{\alpha}_1\cdots\dot{\alpha}_r}
z_{\alpha_1}\cdots z_{\alpha_k}\bar{z}_{\dot{\alpha}_1}\cdots
\bar{z}_{\dot{\alpha}_r}\label{SF}\\
(\alpha_i,\dot{\alpha}_i=0,1),\nonumber
\end{gather}
where the numbers 
$a^{\alpha_1\cdots\alpha_k\dot{\alpha}_1\cdots\dot{\alpha}_r}$
are unaffected at the permutations of indices. 
The expressions (\ref{SF}) can be understood as {\it functions on the
Lorentz group}.\index{function!on the Lorentz group}
When the coefficients
$a^{\alpha_1\cdots\alpha_k\dot{\alpha}_1\cdots\dot{\alpha}_r}$ in
(\ref{SF}) are depend on the variables $x_i\in T_4$ ($i=1,2,3,4$),
we come to the Biedenharn type functions \cite{BBTD88}:
\begin{gather}
p(x,z,\bar{z})=\sum_{\substack{(\alpha_1,\ldots,\alpha_k)\\
(\dot{\alpha}_1,\ldots,\dot{\alpha}_r)}}\frac{1}{k!\,r!}
a^{\alpha_1\cdots\alpha_k\dot{\alpha}_1\cdots\dot{\alpha}_r}(x)
z_{\alpha_1}\cdots z_{\alpha_k}\bar{z}_{\dot{\alpha}_1}\cdots
\bar{z}_{\dot{\alpha}_r}.\label{SF2}\\
(\alpha_i,\dot{\alpha}_i=0,1)\nonumber
\end{gather}
The functions (\ref{SF2}) should be considered as {\it the functions on
the Poincar\'{e} group}.
Some applications of these functions contained
in Refs. 51,45. Representations of 
the Poincar\'{e} group
$SL(2,\C)\odot T(4)$ are realized via the functions (\ref{SF2}).

Let $\cL(\balpha)$ be a Lagrangian on the group manifold $\cM_{10}$ of the
Poincar\'{e} group (in other words, $\cL(\balpha)$ is a 10-dimensional
point function), where $\balpha$ is the parameter set of this group.
Then an integral for $\cL(\balpha)$ on some 10-dimensional volume $\Omega$
of the group manifold we will call {\it an action on the Poincar\'{e}
group}:
\[
A=\int\limits_\Omega d\balpha\cL(\balpha),
\]
where $d\balpha$ is a Haar measure\footnote{The invariant measure $d\balpha$
on the Poincar\'{e} group can be factorized as $d\balpha=dxd\fg$, where
$d\fg$ is a Haar measure on the Lorentz group.}
on the group $\cP$.

Let $\psi(\balpha)$ be a function on the group manifold $\cM_{10}$ (now it is
sufficient to assume that $\psi(\balpha)$ is a square integrable function
on the Poincar\'{e} group) and let
\begin{equation}\label{ELE}
\frac{\partial\cL}{\partial\psi}-\frac{\partial}{\partial\balpha}
\frac{\partial\cL}{\partial\frac{\partial\psi}{\partial\balpha}}=0
\end{equation}
be Euler-Lagrange equations on $\cM_{10}$ (more precisely speaking, the equations
(\ref{ELE}) act on the tangent bundle 
$T\cM_{10}=\underset{\balpha\in\cM_{10}}{\cup}T_{\balpha}\cM_{10}$ 
of the manifold $\cM_{10}$,
see Ref. 52). Let us introduce a Lagrangian $\cL(\balpha)$ depending on
the field function $\psi(\balpha)$ as follows:
\[
\cL(\balpha)=-\frac{1}{2}\left(\psi^\ast(\balpha)B_\mu
\frac{\partial\psi(\balpha)}{\partial\balpha_\mu}-
\frac{\partial\psi^\ast(\balpha)}{\partial\balpha_\mu}B_\mu\psi(\balpha)\right)
-\kappa\psi^\ast(\balpha)B_{11}\psi(\balpha),
\]
where $B_\nu$ ($\nu=1,2,\ldots,10$) are square matrices. The number of
rows and columns in these matrices is equal to the number of components
of $\psi(\balpha)$, $\kappa$ is a non-null real constant.

Further, if $B_{11}$ is non-singular, then we can introduce the matrices
\[
\Pi_\mu=B^{-1}_{11}B_\mu,\quad \mu=1,2,\ldots,10,
\]
and represent the Lagrangian $\cL(\balpha)$ in the form
\[
\cL(\balpha)=-\frac{1}{2}\left(\overline{\psi}(\balpha)\Pi_\mu
\frac{\partial\psi(\balpha)}{\partial\balpha_\mu}-
\frac{\overline{\psi}(\balpha)}{\partial\balpha_\mu}\Pi_\mu
\psi(\balpha)\right)-\kappa\overline{\psi}(\balpha)\psi(\balpha),
\]
where
\[
\overline{\psi}(\balpha)=\psi^\ast(\balpha)B_{11}.
\]
In case of the massless field $(j,0)\oplus(0,j)$ we will consider on the
group manifold $\cM_{10}$ a Lagrangian of the form
\begin{equation}\label{Lagrange}
\cL(\balpha)=-\frac{1}{2}\left(\overline{\psi}(\balpha)\Pi_\mu
\frac{\partial\psi(\balpha)}{\partial\balpha_\mu}-
\frac{\overline{\psi}(\balpha)}{\partial\balpha_\mu}\Pi_\mu
\psi(\balpha)\right).
\end{equation}

As a direct consequence of (\ref{SF2}), the relativistic wavefunction
$\psi(\balpha)$ on the group manifold $\cM_{10}$ is represented by a following
factorization
\begin{equation}\label{WF}
\psi(\balpha)=\psi(x)\psi(\fg)=\psi(x_1,x_2,x_3,x_4)
\psi(\varphi,\epsilon,\theta,\tau,\phi,\varepsilon),
\end{equation}
where $\psi(x_i)$ is a function depending on the parameters of the subgroup
$T_4$, $x_i\in T_4$ ($i=1,\ldots 4$), and $\psi(\fg)$ is a function on the
Lorentz group, where six parameters of this group are defined by the
Euler angles $\varphi$, $\epsilon$, $\theta$, $\tau$, $\phi$, $\varepsilon$
which compose complex angles of the form $\varphi^c=\varphi-i\epsilon$,
$\theta^c=\theta-i\tau$, $\phi^c=\phi-i\varepsilon$ (see Appendix).
\section{The field $(1,0)\oplus(0,1)$}
Before we proceed to start this section let us repeat that the Dirac-like
form of Maxwell equations, considered by many authors, should be derived
from a Dirac-like Lagrangian. It is one of the main assumptions which we
will prove in this paper. Let us rewrite (\ref{Lagrange}) in the form
\begin{multline}\label{Lagrange2}
\cL(\balpha)=-\frac{1}{2}\left(\overline{\psi}(\balpha)\Gamma_\mu
\frac{\partial\psi(\balpha)}{\partial x_\mu}-
\frac{\partial\overline{\psi}(\balpha)}{\partial x_\mu}
\Gamma_\mu\psi(\balpha)\right)-\\
-\frac{1}{2}\left(\overline{\psi}(\balpha)\Upsilon_\nu
\frac{\partial\psi(\balpha)}{\partial\fg_\nu}-
\frac{\partial\overline{\psi}(\balpha)}{\partial\fg_\nu}
\Upsilon_\nu\psi(\balpha)\right),
\end{multline}
where $\psi(\balpha)=\psi(x)\psi(\fg)$ ($\mu=0,1,2,3,\;\nu=1,\ldots,6$),
and
\[
\Gamma_0=\begin{pmatrix}
0 & I\\
I & 0
\end{pmatrix},\;\;\Gamma_1=\begin{pmatrix}
0 & -\alpha_1\\
\alpha_1 & 0
\end{pmatrix},\;\;\Gamma_2=\begin{pmatrix}
0 & -\alpha_2\\
\alpha_2 & 0
\end{pmatrix},\;\;\Gamma_3=\begin{pmatrix}
0 & -\alpha_3\\
\alpha_3 & 0
\end{pmatrix},
\]
\begin{equation}\label{Upsilon1}
\Upsilon_1=\begin{pmatrix}
0 & \Lambda^\ast_1\\
\Lambda_1 & 0
\end{pmatrix},\quad\Upsilon_2=\begin{pmatrix}
0 & \Lambda^\ast_2\\
\Lambda_2 & 0
\end{pmatrix},\quad\Upsilon_3=\begin{pmatrix}
0 & \Lambda^\ast_3\\
\Lambda_3 & 0
\end{pmatrix},
\end{equation}
\begin{equation}\label{Upsilon2}
\Upsilon_4=\begin{pmatrix}
0 & i\Lambda^\ast_1\\
i\Lambda_1 & 0
\end{pmatrix},\quad\Upsilon_5=\begin{pmatrix}
0 & i\Lambda^\ast_2\\
i\Lambda_2 & 0
\end{pmatrix},\quad\Upsilon_6=\begin{pmatrix}
0 & i\Lambda^\ast_3\\
i\Lambda_3 & 0
\end{pmatrix},
\end{equation}
where
\begin{equation}\label{Alpha}
\alpha_1=\begin{pmatrix}
0 & 0 & 0\\
0 & 0 & i\\
0 & -i& 0
\end{pmatrix},\quad\alpha_2=\begin{pmatrix}
0 & 0 & -i\\
0 & 0 & 0\\
i & 0 & 0
\end{pmatrix},\quad\alpha_3=\begin{pmatrix}
0 & i & 0\\
-i& 0 & 0\\
0 & 0 & 0
\end{pmatrix},
\end{equation}
\begin{equation}\label{Lambda}
\Lambda_1=\frac{c_{11}}{\sqrt{2}}\begin{pmatrix}
0 & 1 & 0\\
1 & 0 & 0\\
0 & 1 & 0
\end{pmatrix},\quad\Lambda_2=\frac{c_{11}}{\sqrt{2}}\begin{pmatrix}
0 & -i & 0\\
i & 0 & -i\\
0 & i & 0
\end{pmatrix},\quad\Lambda_3=c_{11}\begin{pmatrix}
1 & 0 & 0\\
0 & 0 & 0\\
0 & 0 &-1
\end{pmatrix}.
\end{equation}
Varying independently $\psi(x)$ and $\overline{\psi}(x)$ in the Lagrangian
(\ref{Lagrange2}), we come to the following equations:
\begin{eqnarray}
\Gamma_\mu\frac{\partial\psi(x)}{\partial x_\mu}&=&0,\label{Real}\\
\Gamma^T_\mu\frac{\partial\overline{\psi}(x)}{\partial x_\mu}&=&0.\label{Anti}
\end{eqnarray}
The equation (\ref{Real}) can be written as follows:
\begin{equation}\label{ME}
\left[\frac{i\hbar}{c}\frac{\partial}{\partial t}\begin{pmatrix}
0 & I\\
I & 0
\end{pmatrix}-i\hbar\frac{\partial}{\partial\bx}\begin{pmatrix}
0 & -\alpha_i\\
\alpha_i & 0
\end{pmatrix}\right]\begin{pmatrix}
\psi(x)\\
\psi^\ast(x)
\end{pmatrix}=0,
\end{equation}
where
\[
\begin{pmatrix}
\psi(x)\\
\psi^\ast(x)
\end{pmatrix}=\begin{pmatrix}
\bE-i\bB\\
\bE+i\bB
\end{pmatrix}=\begin{pmatrix}
E_1-iB_1\\
E_2-iB_2\\
E_3-iE_3\\
E_1+iB_1\\
E_2+iB_2\\
E_3+iB_3
\end{pmatrix}.
\]
From the equation (\ref{ME}) it follows that
\begin{eqnarray}
&&\left(\frac{i\hbar}{c}\frac{\partial}{\partial t}-
i\hbar\alpha_i\frac{\partial}{\partial\bx}\right)\psi(x)=0,\label{ME1}\\
&&\left(\frac{i\hbar}{c}\frac{\partial}{\partial t}+
i\hbar\alpha_i\frac{\partial}{\partial\bx}\right)\psi^\ast(x)=0.\label{ME2}
\end{eqnarray}
The latter equations with allowance for transversality conditions
($\bp\cdot\psi=0$, $\bp\cdot\psi^\ast=0$) coincide with the Maxwell equations.
Indeed, taking into account that 
$(\bp\cdot\alpha)\psi=\hbar\nabla\times\psi$, we obtain
\begin{eqnarray}
\frac{i\hbar}{c}\frac{\partial\psi}{\partial t}&=&-\hbar\nabla\times\psi,
\label{Tr1}\\
-i\hbar\nabla\cdot\psi&=&0.\label{Tr1'}
\end{eqnarray}
Whence
\begin{eqnarray}
\nabla\times(\bE-i\bB)&=&-\frac{i}{c}\frac{\partial(\bE-i\bB)}{\partial t},
\nonumber\\
\nabla\cdot(\bE-i\bB)&=&0\nonumber
\end{eqnarray}
(the constant $\hbar$ is cancelled). Separating the real and imaginary
parts, we obtain Maxwell equations
\begin{eqnarray}
\nabla\times\bE&=&-\frac{1}{c}\frac{\partial\bB}{\partial t},\nonumber\\
\nabla\times\bB&=&\frac{1}{c}\frac{\partial\bE}{\partial t},\nonumber\\
\nabla\cdot\bE&=&0,\nonumber\\
\nabla\cdot\bB&=&0.\nonumber
\end{eqnarray}
It is easy to verify that we come again to Maxwell equations starting from
the equations
\begin{eqnarray}
\left(\frac{i\hbar}{c}\frac{\partial}{\partial t}+
i\hbar\alpha_i\frac{\partial}{\partial\bx}\right)\psi^\ast(x)&=&0,\label{Tr2}\\
-i\hbar\nabla\cdot\psi^\ast(x)&=&0.\label{Tr2'}
\end{eqnarray}
In spite of the fact that equations (\ref{ME1}) and (\ref{ME2}) give rise
to the same Maxwell equations, the physical interpretation of these equations
is different (see Refs. 28,32). Namely, 
the equations (\ref{ME1}) and (\ref{ME2})
are equations with negative and positive 
helicity, 
respectively\footnote{It is interesting to note a correspondence between
photon helicity states and a complexification of the group $SU(2)$
presented by a local isomorphism $SU(2)\otimes SU(2)\simeq SL(2,\C)$.
As known, a root subgroup of a semisimple Lie group $O_4$
(a rotation group of the 4-dimensional space) is a normal divisor of $O_4$.
For that reason the 6-parameter group $O_4$ is semisimple, and is
represented by a direct product of the two 3-parameter unimodular groups.
By analogy with the group $O_4$, a double covering $SL(2,\C)$ of the
proper orthochronous Lorentz group (a rotation group of the
4-dimensional spacetime continuum) is semisimple, and is represented by
a direct product of the two 3-parameter special unimodular groups,
$SL(2,\C)\simeq SU(2)\otimes SU(2)$. An explicit form of this isomorphism
can be obtained by means of a complexification of the group $SU(2)$,
that is, $SL(2,\C)\simeq\mbox{\sf complex}(SU(2))\simeq
SU(2)\otimes SU(2)$ \cite{Var022}.
Moreover, in the works \cite{AE93,Dvo96}, inspired by Ryder book \cite{Ryd85},
the Lorentz group is represented by a product $SU_R(2)\otimes SU_L(2)$,
and spinors
$
\psi(p^\mu)=(\phi_R(p^\mu),\phi_L(p^\mu))^T
$
are transformed within $(j_1,j_2)\oplus(j_2,j_1)$ representation space. The
components 
$\phi_R(p^\mu)$ and $\phi_L(p^\mu)$ correspond to different helicity
states (right- and left-handed spinors). Hence it follows
an analogy with the photon spin states. Namely, the operators
$\sX=\sJ+i\sK$ and $\sY=\sJ-i\sK$ 
correspond to the right and left polarization states of the
photon, where $\sJ$ and $\sK$ are generators of rotation and Lorentz
boosts, respectively.}.

As usual, the conjugated wavefunction 
$\overline{\psi}(x)=\overset{+}{\psi}(x)\Gamma_0=(\psi(x),\psi^\ast(x))$
corresponds to antiparticle (it is a direct consequence of the
Dirac-like Lagrangian (\ref{Lagrange2}), $\psi(x)$ is a complex
function). Therefore, we come to a very controversial conclusion that
the equations (\ref{Anti}) describes the antiparticle (antiphoton)
and, moreover, hence it follows that there exist the current and
charge for the photon field. At first glance, we come to a drastic
contradiction with the widely accepted fact that the photon is
truly neutral particle. However, it is easy to verify that
equations (\ref{Anti}) give rise to the Maxwell equations also. It means that
the photon coincides with its ``antiparticle". Following to the standard
procedure given in many textbooks, we can define the ``charge" of the
photon by an expression
\begin{equation}\label{Charge}
Q\sim\int d\bx\overline{\psi}\Gamma_0\psi,
\end{equation}
where $\overline{\psi}\Gamma_0\psi=2(\bE^2+\bB^2)$. However, Newton and
Wigner \cite{NW49} showed that for the photon there exist no localized
states. Therefore, the integral in the right side of (\ref{Charge})
presents an indeterminable expression. Since the integral (\ref{Charge})
does not exist in general, then the ``charge" of the photon cannot be
considered as a constant magnitude (as it takes place for the electron
field which has localized states \cite{NW49} and a well-defined
constant charge). In a sense, one can say that the ``charge " of the
photon is equal to the energy $\bE^2+\bB^2$ of $\gamma$-quantum.

We see that the equation (\ref{ME}) gives rise to the two Dirac-like
equations (\ref{ME1}) and (\ref{ME2}) which in combination with the
transversality conditions (\ref{Tr1'}) and (\ref{Tr2'}) are equivalent to
the Maxwell equations. Let us represent solutions of (\ref{ME1}) in 
a plane-wave form
\begin{equation}\label{PW}
\psi(x)=\varepsilon(\bk)\exp[i\hbar^{-1}(\bk\cdot\bx-\omega t)].
\end{equation}
After substitution of (\ref{PW}) into (\ref{ME1}) we come to the following
matrix eigenvalue problem
\[
-c\begin{pmatrix}
0 & ik_3 & -ik_2\\
-ik_3 & 0 & ik_1\\
ik_2 & -ik_1 & 0
\end{pmatrix}\begin{pmatrix}
\varepsilon_1\\
\varepsilon_2\\
\varepsilon_3
\end{pmatrix}=\omega\begin{pmatrix}
\varepsilon_1\\
\varepsilon_2\\
\varepsilon_3
\end{pmatrix}.
\]
It is easy to verify that we come to the same eigenvalue problem starting
from (\ref{ME2}). The secular equation has the solutions
$\omega=\pm c\bk,0$.

Therefore, solutions of (\ref{ME}) in the plane-wave 
approximation are expressed via the functions
\begin{eqnarray}
\psi_\pm(\bk;\bx,t)&=&\lf 2(2\pi)^3\rf^{-\frac{1}{2}}
\begin{pmatrix}
\varepsilon_\pm(\bk)\\
\varepsilon_\pm(\bk)
\end{pmatrix}
\exp[i(\bk\cdot\bx-\omega t)],\nonumber\\
\psi_0(\bk;\bx)&=&\lf 2(2\pi)^3\rf^{-\frac{1}{2}}
\begin{pmatrix}
\varepsilon_0(\bk)\\
\varepsilon_0(\bk)
\end{pmatrix}
\exp[i\bk\cdot\bx]\nonumber
\end{eqnarray}
and the complex conjugate functions $\psi^\ast_+(\bk;\bx,t)$ and
$\psi^\ast_0(\bk;\bx)$ ($\bE+i\bB$) corresponding to positive helicity,
here $\omega=c|\bk|$ and $\varepsilon_\lambda(\bk)$ ($\lambda=\pm,0$)
are the polarization vectors of a photon:
\begin{eqnarray}
\varepsilon_\pm(\bk)&=&\lf 2|\bk|^2(k^2_1+k^2_2)\rf^{-\frac{1}{2}}
\begin{bmatrix}
-k_1k_3\pm ik_2|\bk|\\
-k_2k_3\mp ik_1|\bk|\\
k^2_1+k^2_2
\end{bmatrix},\nonumber\\
\varepsilon_0(\bk)&=&|\bk|^{-1}
\begin{bmatrix}
k_1\\
k_2\\
k_3
\end{bmatrix}.\nonumber
\end{eqnarray}

Varying now $\psi(\fg)$ and $\overline{\psi}(\fg)$ in the Lagrangian
(\ref{Lagrange2}), we come to equations
\begin{eqnarray}
\Upsilon_\nu\frac{\partial\psi(\fg)}{\partial\fg_\nu}&=&0,\nonumber\\
\Upsilon^T_\nu\frac{\partial\overline{\psi}(\fg)}{\partial\fg_\nu}&=&0.
\label{Upsilon3}
\end{eqnarray}
The latter equations are written in the parameters of $SL(2,\C)$. Since an
universal covering $SL(2,\C)$ of the proper orthochronous Lorentz group
is a complexification of the group $SU(2)$ 
(see, for example, Ref. 58),
then it is more convenient to express six parameters of the group
$SL(2,\C)$ via three parameters $a_1$, $a_2$, $a_3$ of $SU(2)$. It is
obviuos that $\fg_1=a_1$, $\fg_2=a_2$, $\fg_3=a_3$, $\fg_4=ia_1$, 
$\fg_5=ia_2$, $\fg_6=ia_3$. Taking into account the structure of 
$\Upsilon_\nu$ given by (\ref{Upsilon1})--(\ref{Upsilon2}), we can rewrite
the first equation from (\ref{Upsilon3}) as follows
\begin{eqnarray}
\sum^3_{k=1}\Lambda_k\frac{\partial\psi}{\partial a_k}-
i\sum^3_{k=1}\Lambda_k\frac{\partial\psi}{\partial a^\ast_k}&=&0,\nonumber\\
\sum^3_{k=1}\Lambda^\ast_k\frac{\partial\dot{\psi}}{\partial\widetilde{a}_k}+
i\sum^3_{k=1}\Lambda^\ast_k\frac{\partial\dot{\psi}}
{\partial\widetilde{a}^\ast_k}&=&0.\label{Complex}
\end{eqnarray}
where $a^\ast_1=-i\fg_4$, $a^\ast_2=-i\fg_5$, $a^\ast_3=-i\fg_6$, and
$\widetilde{a}_k$, $\widetilde{a}^\ast_k$ are parameters corresponding
to dual basis. In essence, equations (\ref{Complex}) are defined
in a three-dimensional complex space $\C^3$. In turn, the space $\C^3$
is isometric to a six-dimensional bivector space $\R^6$ (a parameter
space of the Lorentz group \cite{Pet69}). The bivector space $\R^6$
is a tangent space of the group manifold $\fL_6$ of the Lorentz group,
that is, the group manifold $\fL_6$ in the each its point is equivalent
to the space $\R^6$. Thus, for all $\fg\in\fL_6$ we have
$T_{\fg}\fL_6\simeq\R^6$. Taking into account the explicit 
form\footnote{The matrices (\ref{Lambda}) can be derived at $l=1$ from
the more general expressions (62)--(67)
given in the Ref. 62} of the matrices $\Lambda_i$
given by (\ref{Lambda}), we can rewrite the system (\ref{Complex})
in the following form
\begin{gather}
\frac{\sqrt{2}}{2}\frac{\partial\psi_2}{\partial x_1}-
i\frac{\sqrt{2}}{2}\frac{\partial\psi_2}{\partial x_2}+
\frac{\partial\psi_1}{\partial x_3}-
i\frac{\sqrt{2}}{2}\frac{\partial\psi_2}{\partial x^\ast_1}-
\frac{\sqrt{2}}{2}\frac{\partial\psi_2}{\partial x^\ast_2}-
i\frac{\partial\psi_1}{\partial x^\ast_3}=0,\nonumber\\
\frac{\partial\psi_1}{\partial x_1}+
\frac{\partial\psi_3}{\partial x_1}+
i\frac{\partial\psi_1}{\partial x_2}-
i\frac{\partial\psi_3}{\partial x_2}-
i\frac{\partial\psi_1}{\partial x^\ast_1}-
i\frac{\partial\psi_3}{\partial x^\ast_1}-
\frac{\partial\psi_1}{\partial x^\ast_2}+
\frac{\partial\psi_3}{\partial x^\ast_2}=0,\nonumber\\
\frac{\sqrt{2}}{2}\frac{\partial\psi_2}{\partial x_1}+
i\frac{\sqrt{2}}{2}\frac{\partial\psi_2}{\partial x_2}-
\frac{\partial\psi_3}{\partial x_3}-
i\frac{\sqrt{2}}{2}\frac{\partial\psi_2}{\partial x^\ast_1}+
\frac{\sqrt{2}}{2}\frac{\partial\psi_2}{\partial x^\ast_2}+
i\frac{\partial\psi_3}{\partial x^\ast_3}=0,\nonumber\\
\frac{\sqrt{2}}{2}\frac{\partial\dot{\psi}_2}{\partial\widetilde{x}_1}-
i\frac{\sqrt{2}}{2}\frac{\partial\dot{\psi}_2}{\partial\widetilde{x}_2}+
\frac{\partial\dot{\psi}_1}{\partial\widetilde{x}_3}+
i\frac{\sqrt{2}}{2}\frac{\partial\dot{\psi}_2}{\partial\widetilde{x}^\ast_1}+
\frac{\sqrt{2}}{2}\frac{\partial\dot{\psi}_2}{\partial\widetilde{x}^\ast_2}+
i\frac{\partial\dot{\psi}_1}{\partial\widetilde{x}^\ast_3}=0,\nonumber\\
\frac{\partial\dot{\psi}_1}{\partial\widetilde{x}_1}+
\frac{\partial\dot{\psi}_3}{\partial\widetilde{x}_1}+
i\frac{\partial\dot{\psi}_1}{\partial\widetilde{x}_2}-
i\frac{\partial\dot{\psi}_3}{\partial\widetilde{x}_2}+
i\frac{\partial\dot{\psi}_1}{\partial\widetilde{x}^\ast_1}+
i\frac{\partial\dot{\psi}_3}{\partial\widetilde{x}^\ast_1}-
\frac{\partial\dot{\psi}_1}{\partial\widetilde{x}^\ast_2}+
\frac{\partial\dot{\psi}_3}{\partial\widetilde{x}^\ast_2}=0,\nonumber\\
\frac{\sqrt{2}}{2}\frac{\partial\dot{\psi}_2}{\partial\widetilde{x}_1}+
i\frac{\sqrt{2}}{2}\frac{\partial\dot{\psi}_2}{\partial\widetilde{x}_2}-
\frac{\partial\dot{\psi}_3}{\partial\widetilde{x}_3}+
i\frac{\sqrt{2}}{2}\frac{\partial\dot{\psi}_2}{\partial\widetilde{x}^\ast_1}-
\frac{\sqrt{2}}{2}\frac{\partial\dot{\psi}_2}{\partial\widetilde{x}^\ast_2}-
i\frac{\partial\dot{\psi}_3}{\partial\widetilde{x}^\ast_3}=0,\label{CompM}
\end{gather}
\begin{sloppypar}
A separation of variables in (\ref{CompM}) is realized via the following
factorizations:
\end{sloppypar}
\begin{eqnarray}
\psi_1&=&\boldsymbol{f}^l_{1,1}(r)
\fM_l^{1}(\varphi,\epsilon,\theta,\tau,0,0),\nonumber\\
\psi_2&=&\boldsymbol{f}^l_{1,0}(r)
\fM_l^{0}(0,0,\theta,\tau,0,0),\nonumber\\
\psi_3&=&\boldsymbol{f}^l_{1,-1}(r)
\fM_l^{-1}(\varphi,\epsilon,\theta,\tau,0,0),\nonumber\\
\dot{\psi}_1&=&
\boldsymbol{f}^{\dot{l}}_{1,1}(r^\ast)
\fM_{\dot{l}}^{1}(\varphi,\epsilon,\theta,\tau,0,0),\nonumber\\
\dot{\psi}_2&=&
\boldsymbol{f}^{\dot{l}}_{1,0}(r^\ast)
\fM_{\dot{l}}^{0}(0,0,\theta,\tau,0,0),\nonumber\\
\dot{\psi}_3&=&
\boldsymbol{f}^{\dot{l}}_{1,-1}(r^\ast)
\fM_{\dot{l}}^{-1}(\varphi,\epsilon,\theta,\tau,0,0),\nonumber
\end{eqnarray}
where $\fM_{l}^m(\varphi,\epsilon,\theta,\tau,0,0)$
($\fM_{\dot{l}}^{\dot{m}}(\varphi,\epsilon,\theta,\tau,0,0)$) are
associated hyperspherical functions defined on the surface of the 
two-dimensional complex sphere of the radius $r$, 
$\boldsymbol{f}^{l_0}_{lmk}(r)$ and
$\boldsymbol{f}^{\dot{l}_0}_{\dot{l}\dot{m}\dot{k}}(r^\ast)$ are radial
functions (for more details about two-dimensional complex sphere see
Refs. 60--62)\footnote{Choosing the two-dimensional complex
sphere as an internal spin space, we see that our configuration space
$\cM_{10}=\R^{1,3}\times\fL_6$ reduces to $\cM_8=\R^{1,3}\times\C^2$.
Bacry and Kihlberg \cite{BK69} claimed that the 8-dimensional configuration
space $\cM_8$ is the most suitable for a description of both half-integer
and integer spins. $\cM_8$ is a homogeneous space of the Poincar\'{e}
group. Indeed, the space $\C^2$ is homeomorphic to an extended complex
plane $\C\cup\infty$ which presents an absolute (a set of infinitely
distant points) of a Lobachevskii space $S^{1,2}$. At this point,
the group of fractional linear transformations of the plane
$\C\cup\infty$ is isomorphic to a motion group of $S^{1,2}$. In turn,
the Lobachevskii space $S^{1,2}$ is an absolute of the Minkowski world
$\R^{1,3}$ and, therefore, the group of fractional linear transformations
of $\C\cup\infty$ double covers a rotation group of $\R^{1,3}$, that is,
the Lorentz group. It is not hard to see that the two-dimensional
complex sphere coincides with a well-known 
Penrose's celestial sphere \cite{Pen84}. Further, using a canonical
projection $\pi:\;\C^2_\ast\rightarrow S^2$, where 
$\C^2_\ast=\C^2/\{0,0\}$ and $S^2$ is a two-dimensional real sphere,
we see that the configuration space $\cM_8=\R^{1,3}\times\C^2$ reduces
to $\cM_6=\R^{1,3}\times S^2$. The real two-sphere $S^2$ has a minimal
possible dimension among the homogeneous spaces of the Lorentz group.
For that reason $\cM_6$ is a minimal homogeneous space of the
Poincar\'{e} group (the spacetime translations act trivially on $S^2$).
Field models on the configuration space $\cM_6$ have been considered in
recent works \cite{KLS95,LSS96,Dre97}. In the Ref. 43
Drechsler considered the two-sphere as a ``spin shell"
$S^2_{r=2s}$ of radius $r=2s$, where $s=0,\frac{1}{2},1,\frac{3}{2},\ldots$.}. 

Repeating for the case $l=1$ all the transformations presented for the
general relativistically invariant system \cite{Var031}, 
we come to the following system
(the system (126) in the Ref. 62 at $l=1$):
\begin{gather}
2\frac{d\boldsymbol{f}^l_{1,1}(r)}{dr}-
\frac{1}{r}\boldsymbol{f}^l_{1,1}(r)-
\frac{\sqrt{2l(l+1)}}{r}\boldsymbol{f}^l_{1,0}(r)=0,\nonumber\\
-\frac{\sqrt{2l(l+1)}}{r}\boldsymbol{f}^l_{1,-1}(r)+
\frac{\sqrt{2l(l+1)}}{r}\boldsymbol{f}^l_{1,1}(r)=0,\nonumber\\
-2\frac{d\boldsymbol{f}^l_{1,-1}(r)}{dr}+\frac{1}{r}\boldsymbol{f}^l_{1,-1}(r)+
\frac{\sqrt{2l(l+1)}}{r}\boldsymbol{f}^l_{1,0}(r)=0,\nonumber\\
2\frac{d\boldsymbol{f}^{\dot{l}}_{1,1}(r^\ast)}{dr^\ast}-\frac{1}{r^\ast}
\boldsymbol{f}^{\dot{l}}_{1,1}(r^\ast)-\frac{\sqrt{2\dot{l}(\dot{l}+1)}}{r^\ast}
\boldsymbol{f}^{\dot{l}}_{1,0}(r^\ast)=0,\nonumber\\
-\frac{i\sqrt{2\dot{l}(\dot{l}+1)}}{r^\ast}
\boldsymbol{f}^{\dot{l}}_{1,-1}(r^\ast)+\frac{\sqrt{2\dot{l}(\dot{l}+1)}}{r^\ast}
\boldsymbol{f}^{\dot{l}}_{1,1}(r^\ast)=0,\nonumber\\
-2\frac{d\boldsymbol{f}^{\dot{l}}_{1,-1}(r^\ast)}{dr^\ast}+\frac{1}{r^\ast}
\boldsymbol{f}^{\dot{l}}_{1,-1}(r^\ast)+\frac{\sqrt{2\dot{l}(\dot{l}+1)}}{r^\ast}
\boldsymbol{f}^{\dot{l}}_{1,0}(r^\ast)=0,\label{RFM}
\end{gather}
From the second and fifth equations it follows that
$\boldsymbol{f}^l_{1,-1}(r)=\boldsymbol{f}^l_{1,1}(r)$
and $\boldsymbol{f}^{\dot{l}}_{1,-1}(r^\ast)=
\boldsymbol{f}^{\dot{l}}_{1,1}(r^\ast)$. 
Taking into account these relations we can rewrite
the system (\ref{RFM}) as follows
\begin{eqnarray}
2\frac{d\boldsymbol{f}^l_{1,1}(r)}{dr}\;\;\;-\;\;
\frac{1}{r}\boldsymbol{f}^l_{1,1}(r)\;\;-\;\;
\frac{\sqrt{2l(l+1)}}{r}\boldsymbol{f}^l_{1,0}(r)
&=&0,\nonumber\\
-2\frac{d\boldsymbol{f}^l_{1,-1}(r)}{dr}\;\;+\,\,
\frac{1}{r}\boldsymbol{f}^l_{1,-1}(r)\;+\;
\frac{\sqrt{2l(l+1)}}{r}\boldsymbol{f}^l_{1,0}(r)
&=&0,\nonumber\\
2\frac{d\boldsymbol{f}^{\dot{l}}_{1,1}(r^\ast)}{dr^\ast}\,-\;\frac{1}{r^\ast}
\boldsymbol{f}^{\dot{l}}_{1,1}(r^\ast)\;-\;\frac{\sqrt{2\dot{l}(\dot{l}+1)}}{r^\ast}
\boldsymbol{f}^{\dot{l}}_{1,0}(r^\ast)&=&0,\nonumber\\
-2\frac{d\boldsymbol{f}^{\dot{l}}_{1,-1}(r^\ast)}{dr^\ast}+\frac{1}{r^\ast}
\boldsymbol{f}^{\dot{l}}_{1,-1}(r^\ast)+\frac{\sqrt{2\dot{l}(\dot{l}+1)}}{r^\ast}
\boldsymbol{f}^{\dot{l}}_{1,0}(r^\ast)&=&0,\label{RFM2}
\end{eqnarray}
It is easy to see that the first equation is equivalent to the second,
and third equation is equivalent to the fourth. Thus, we come to the
following inhomogeneous differential equations of the first order:
\begin{eqnarray}
&&2r\frac{d\boldsymbol{f}^l_{1,1}(r)}{dr}-\boldsymbol{f}^l_{1,1}(r)-
\sqrt{2l(l+1)}\boldsymbol{f}^l_{1,0}(r)=0,\nonumber\\
&&2r^\ast\frac{d\boldsymbol{f}^{\dot{l}}_{1,1}(r^\ast)}{dr^\ast}-
\boldsymbol{f}^{\dot{l}}_{1,1}(r^\ast)-
\sqrt{2\dot{l}(\dot{l}+1)}\boldsymbol{f}^{\dot{l}}_{1,0}(r^\ast)=0,\nonumber
\end{eqnarray}
where the functions $\boldsymbol{f}^l_{1,0}(r)$ and
$\boldsymbol{f}^{\dot{l}}_{1,0}(r^\ast)$ are understood as inhomogeneous
parts. Solutions of these equations are expressed via the elementary functions:
\begin{eqnarray}
\boldsymbol{f}^l_{1,1}(r)&=&C\sqrt{r}+\sqrt{2l(l+1)}r,\nonumber\\
\boldsymbol{f}^{\dot{l}}_{1,1}(r^\ast)&=&\dot{C}\sqrt{r^\ast}+
\sqrt{2\dot{l}(\dot{l}+1)}r^\ast.\nonumber
\end{eqnarray}
Therefore, solutions of the radial part have the form:
\begin{eqnarray}
&&\boldsymbol{f}^l_{1,1}(r)=\boldsymbol{f}^l_{1,-1}(r)=
C\sqrt{r}+\sqrt{2l(l+1)}r,\nonumber\\
&&\boldsymbol{f}^l_{1,0}(r)=\sqrt{2l(l+1)}r,\nonumber\\
&&\boldsymbol{f}^{\dot{l}}_{1,1}(r^\ast)=
\boldsymbol{f}^{\dot{l}}_{1,-1}(r^\ast)=
\dot{C}\sqrt{r^\ast}+\sqrt{2\dot{l}(\dot{l}+1)}r^\ast,\nonumber\\
&&\boldsymbol{f}^{\dot{l}}_{1,0}(r^\ast)=
\sqrt{2\dot{l}(\dot{l}+1)}r^\ast.
\nonumber
\end{eqnarray}
In such a way, solutions of the $SL(2,\C)$--field equations 
(\ref{CompM}) are
\begin{eqnarray}
\psi_1(r,\varphi^c,\theta^c)&=&\boldsymbol{f}^l_{1,1}(r)
\fM_l^{1}(\varphi,\epsilon,\theta,\tau,0,0),\nonumber\\
\psi_2(r,\varphi^c,\theta^c)&=&\boldsymbol{f}^l_{1,0}(r)
\fM_l^{0}(0,0,\theta,\tau,0,0),\nonumber\\
\psi_3(r,\varphi^c,\theta^c)&=&\boldsymbol{f}^l_{1,-1}(r)
\fM_l^{-1}(\varphi,\epsilon,\theta,\tau,0,0),\nonumber\\
\dot{\psi}_1(r^\ast,\dot{\varphi}^c,\dot{\theta}^c)&=&
\boldsymbol{f}^{\dot{l}}_{1,1}(r^\ast)
\fM_{\dot{l}}^{1}(\varphi,\epsilon,\theta,\tau,0,0),\nonumber\\
\dot{\psi}_2(r^\ast,\dot{\varphi}^c,\dot{\theta}^c)&=&
\boldsymbol{f}^{\dot{l}}_{1,0}(r^\ast)
\fM_{\dot{l}}^{0}(0,0,\theta,\tau,0,0),\nonumber\\
\dot{\psi}_3(r^\ast,\dot{\varphi}^c,\dot{\theta}^c)&=&
\boldsymbol{f}^{\dot{l}}_{1,-1}(r^\ast)
\fM_{\dot{l}}^{-1}(\varphi,\epsilon,\theta,\tau,0,0),\nonumber
\end{eqnarray}
where
\begin{eqnarray}
&&l=1,\;2,\;3,\;\ldots\nonumber\\
&&\dot{l}=1,\;2,\;3,\;\ldots\nonumber
\end{eqnarray}
\[
\fM_l^{\pm 1}(\varphi,\epsilon,\theta,\tau,0,0)=
e^{\mp(\epsilon+i\varphi)}Z_l^{\pm 1}(\theta,\tau),
\]
\begin{multline}
Z_l^{\pm 1}(\theta,\tau)=\cos^{2l}\frac{\theta}{2}
\ch^{2l}\frac{\tau}{2}\sum^l_{k=-l}i^{\pm 1-k}
\tg^{\pm 1-k}\frac{\theta}{2}\tnh^{-k}\frac{\tau}{2}\times\\
\hypergeom{2}{1}{\pm 1-l+1,1-l-k}{\pm 1-k+1}
{i^2\tg^2\frac{\theta}{2}}
\hypergeom{2}{1}{-l+1,1-l-k}{-k+1}{\tnh^2\frac{\tau}{2}},\nonumber
\end{multline}
\[
\fM_l^{0}(0,0,\theta,\tau,0,0)=
Z_l^{0}(\theta,\tau),
\]
\begin{multline}
Z_l^{0}(\theta,\tau)=\cos^{2l}\frac{\theta}{2}
\ch^{2l}\frac{\tau}{2}\sum^l_{k=-l}i^{-k}
\tg^{-k}\frac{\theta}{2}\tnh^{-k}\frac{\tau}{2}\times\\
\hypergeom{2}{1}{-l+1,1-l-k}{-k+1}
{i^2\tg^2\frac{\theta}{2}}
\hypergeom{2}{1}{-l+1,1-l-k}{-k+1}{\tnh^2\frac{\tau}{2}},\nonumber
\end{multline}
\[
\fM_{\dot{l}}^{\pm 1,}(\varphi,\epsilon,\theta,\tau,0,0)=
e^{\mp(\epsilon-i\varphi)}
Z_{\dot{l}}^{\pm 1}(\theta,\tau),
\]
\begin{multline}
Z_{\dot{l}}^{\pm 1}(\theta,\tau)=
\cos^{2\dot{l}}\frac{\theta}{2}
\ch^{2\dot{l}}\frac{\tau}{2}
\sum^{\dot{l}}_{\dot{k}=-\dot{l}}i^{\pm 1-\dot{k}}
\tg^{\pm 1-\dot{k}}\frac{\theta}{2}
\tnh^{-\dot{k}}\frac{\tau}{2}\times\\
\hypergeom{2}{1}{\pm 1-\dot{l}+1,1-\dot{l}-\dot{k}}
{\pm 1-\dot{k}+1}
{i^2\tg^2\frac{\theta}{2}}
\hypergeom{2}{1}{-\dot{l}+1,1-\dot{l}-\dot{k}}
{-\dot{k}+1}{\tnh^2\frac{\tau}{2}},\nonumber
\end{multline}
\[
\fM_{\dot{l}}^{0}(0,0,\theta,\tau,0,0)=
Z_{\dot{l}}^{0}(\theta,\tau),
\]
\begin{multline}
Z_{\dot{l}}^{0}(\theta,\tau)=
\cos^{2\dot{l}}\frac{\theta}{2}
\ch^{2\dot{l}}\frac{\tau}{2}
\sum^{\dot{l}}_{\dot{k}=-\dot{l}}i^{-\dot{k}}
\tg^{-\dot{k}}\frac{\theta}{2}
\tnh^{-\dot{k}}\frac{\tau}{2}\times\\
\hypergeom{2}{1}{-\dot{l}+1,1-\dot{l}-\dot{k}}
{-\dot{k}+1}
{i^2\tg^2\frac{\theta}{2}}
\hypergeom{2}{1}{-\dot{l}+1,1-\dot{l}-\dot{k}}
{-\dot{k}+1}{\tnh^2\frac{\tau}{2}},\nonumber
\end{multline}
where $\fM_l^0(0,0,\theta,\tau,0,0)$ ($Z_l^0(\theta,\tau)$) are zonal
hyperspherical functions (see Ref. 48).

Therefore, in accordance with the factorization (\ref{WF}) an explicit
form of the relativistic wavefunction $\psi(\balpha)=\psi(x)\psi(\fg)$
on the Poincar\'{e} group in the case of $(1,0)\oplus(0,1)$ representation
(Maxwell field) is defined by the following expressions (complete set)
\begin{multline}
\psi_1(\balpha)=\psi_+(\bk;\bx,t)\psi_1(\fg)=\\
\lf 2(2\pi)^3\rf^{-\frac{1}{2}}
\begin{pmatrix}\varepsilon_+(\bk)\\
\varepsilon_+(\bk)\end{pmatrix}
\exp[i(\bk\cdot\bx-\omega t)]\boldsymbol{f}^l_{1,1}(r)
\fM^1_l(\varphi,\epsilon,\theta,\tau,0,0),\nonumber
\end{multline}
\[
\psi_0(\balpha)=\psi_0(\bk;\bx)\psi_0(\fg)=\lf 2(2\pi)^3\rf^{-\frac{1}{2}}
\begin{pmatrix}\varepsilon_0(\bk)\\
\varepsilon_0(\bk)\end{pmatrix}
\exp[i\bk\cdot\bx]\boldsymbol{f}^l_{1,0}(r)
\fM_l(0,0,\theta,\tau,0,0),
\]
\begin{multline}
\psi_{-1}(\balpha)=\psi_-(\bk;\bx,t)\psi_{-1}(\fg)=\\
\lf 2(2\pi)^3\rf^{-\frac{1}{2}}
\begin{pmatrix}\varepsilon_-(\bk)\\
\varepsilon_-(\bk)\end{pmatrix}
\exp[i(\bk\cdot\bx-\omega t)]\boldsymbol{f}^l_{1,-1}(r)
\fM^{-1}_l(\varphi,\epsilon,\theta,\tau,0,0),\nonumber
\end{multline}
\begin{multline}
\dot{\psi}_1(\balpha)=\psi^\ast_+(\bk;\bx,t)\dot{\psi}_1(\fg)=\\
\lf 2(2\pi)^3\rf^{-\frac{1}{2}}
\begin{pmatrix}\varepsilon^\ast_+(\bk)\\
\varepsilon^\ast_+(\bk)\end{pmatrix}
\exp[-i(\bk\cdot\bx-\omega t)]\boldsymbol{f}^l_{1,1}(r^\ast)
\fM^1_{\dot{l}}(\varphi,\epsilon,\theta,\tau,0,0),\nonumber
\end{multline}
\[
\dot{\psi}_0(\balpha)=\psi^\ast_0(\bk;\bx)\dot{\psi}_0(\fg)=
\lf 2(2\pi)^3\rf^{-\frac{1}{2}}
\begin{pmatrix}\varepsilon^\ast_0(\bk)\\
\varepsilon^\ast_0(\bk)\end{pmatrix}
\exp[-i\bk\cdot\bx]\boldsymbol{f}^l_{1,0}(r^\ast)
\fM_{\dot{l}}(0,0,\theta,\tau,0,0),
\]
\begin{multline}
\dot{\psi}_{-1}(\balpha)=\psi^\ast_-(\bk;\bx,t)\dot{\psi}_{-1}(\fg)=\\
\lf 2(2\pi)^3\rf^{-\frac{1}{2}}
\begin{pmatrix}\varepsilon^\ast_-(\bk)\\
\varepsilon^\ast_-(\bk)\end{pmatrix}
\exp[-i(\bk\cdot\bx-\omega t)]\boldsymbol{f}^l_{1,-1}(r^\ast)
\fM^{-1}_{\dot{l}}(\varphi,\epsilon,\theta,\tau,0,0),\label{SME}
\end{multline}
The set (\ref{SME}) consists of the transverse solutions $\psi_{\pm 1}(\balpha)$
(positive energy), $\dot{\psi}_{\pm 1}(\balpha)$ (negative energy) and
the zero-eigenvalue (longitudinal) solutions $\psi_0(\balpha)$ and
$\dot{\psi}_0(\balpha)$. The negative energy solutions 
$\dot{\psi}_{\pm 1}(\balpha)$ should be omitted, since photons have no
antiparticles.
The longitudinal solutions $\psi_0(\balpha)$ and
$\dot{\psi}_0(\balpha)$ do not contribute to a real photon due to their
transversality conditions (\ref{Tr1'}) and (\ref{Tr2'}). Thus, any real
photon should be described by only $\psi_{\pm 1}(\balpha)$:
\begin{multline}
\psi_{\pm 1}(\balpha)=\psi_{\pm}(\bk;\bx,t)\psi_{\pm 1}(\fg)=\\
\lf 2(2\pi)^3\rf^{-\frac{1}{2}}
\begin{pmatrix}\varepsilon_\pm(\bk)\\
\varepsilon_\pm(\bk)\end{pmatrix}
\exp[i(\bk\cdot\bx-\omega t)]\boldsymbol{f}^l_{1,\pm 1}(r)
\fM^{\pm 1}_l(\varphi,\epsilon,\theta,\tau,0,0),\nonumber
\end{multline}
In such a way, we obtain a solution set defining the Maxwell field
$(1,0)\oplus (0,1)$ on the Poincar\'{e} group (or, equally, on the group
manifold $\cM_8$). It should be noted that obtained previously solutions
for the field $(1/2,0)\oplus(0,1/2)$ (Dirac field) \cite{Var035} have the
analogous mathematical structure, that is, they are the functions on the
Poincar\'{e} group. This circumstance allows us to consider the fields
$(1/2,0)\oplus(0,1/2)$ and $(1,0)\oplus(0,1)$ on an equal footing, from
the one group theoretical viewpoint. The following step is a definition of
field operators in the obtained solutions for the Dirac and Maxwell fields
via harmonic analysis on the Poincar\'{e} group. A construction of
quantum electrodynamics in terms of so-defined field operators will be
studied in a separate work.
\section*{Appendix. Complex two-sphere and hyperspherical functions}
\setcounter{equation}{0}
\renewcommand{\theequation}{A.\arabic{equation}}
Let us construct in $\C^3$ a two--dimensional complex sphere from the
quantities $z_k=x_k+iy_k$, $\overset{\ast}{z}_k=x_k-iy_k$
as follows
\begin{equation}\label{CS}
\bz^2=z^2_1+z^2_2+z^2_3=\bx^2-\by^2+2i\bx\by=r^2
\end{equation}
and its complex conjugate (dual) sphere
\begin{equation}\label{DS}
\overset{\ast}{\bz}{}^2=\overset{\ast}{z}_1{}^2+\overset{\ast}{z}_2{}^2+
\overset{\ast}{z}_3{}^2=\bx^2-\by^2-2i\bx\by=\overset{\ast}{r}{}^2.
\end{equation}
It is well-known that both quantities $\bx^2-\by^2$, $\bx\by$ are
invariant with respect to the Lorentz transformations, since a surface of
the complex sphere is invariant 
(Casimir operators of the Lorentz group are
constructed from such quantities, see also (\ref{KO})).
Moreover, since the real and imaginary parts of the complex two-sphere
transform like the electric and magnetic fields, respectively,
the invariance of $\bz^2\sim(\bE+i\bB)^2$ under proper Lorentz
transformations is evident.

The group $SL(2,\C)$ of all complex matrices
\[\ar
\begin{pmatrix}
\alpha & \beta\\
\gamma & \delta
\end{pmatrix}
\]
of 2-nd order with the determinant $\alpha\delta-\gamma\beta=1$, is
a {\it complexification} of the group $SU(2)$. The group $SU(2)$ is one of
the real forms of $SL(2,\C)$. The transition from $SU(2)$ to $SL(2,\C)$
is realized via the complexification of three real parameters
$\varphi,\,\theta,\,\psi$ (Euler angles). Let $\theta^c=\theta-i\tau$,
$\varphi^c=\varphi-i\epsilon$, $\psi^c=\psi-i\varepsilon$ be complex
Euler angles, where
\begin{equation}\label{CEA}
{\renewcommand{\arraystretch}{1.05}
\begin{array}{ccccc}
0 &\leq&\re\theta^c=\theta& \leq& \pi,\\
0 &\leq&\re\varphi^c=\varphi& <&2\pi,\\
-2\pi&\leq&\re\psi^c=\psi&<&2\pi,
\end{array}\quad\quad
\begin{array}{ccccc}
-\infty &<&\im\theta^c=\tau&<&+\infty,\\
-\infty&<&\im\varphi^c=\epsilon&<&+\infty,\\
-\infty&<&\im\psi^c=\varepsilon&<&+\infty.
\end{array}}
\end{equation}
As known, for the Lorentz group there are two independent 
Casimir operators
\begin{eqnarray}
\sX^2&=&\sX^2_1+\sX^2_2+\sX^2_3=\frac{1}{4}(\sA^2-\sB^2+2i\sA\sB),\nonumber\\
\sY^2&=&\sY^2_1+\sY^2_2+\sY^2_3=
\frac{1}{4}(\widetilde{\sA}^2-\widetilde{\sB}^2-
2i\widetilde{\sA}\widetilde{\sB}).\label{KO}
\end{eqnarray}
Using the parameters (\ref{CEA}), we obtain 
for the Casimir operators the following expressions
\begin{eqnarray}
\sX^2&=&\frac{\partial^2}{\partial\theta^c{}^2}+
\ctg\theta^c\frac{\partial}{\partial\theta^c}+\frac{1}{\sin^2\theta^c}\left[
\frac{\partial^2}{\partial\varphi^c{}^2}-
2\cos\theta^c\frac{\partial}{\partial\varphi^c}
\frac{\partial}{\partial\psi^c}+
\frac{\partial^2}{\partial\psi^c{}^2}\right],\nonumber\\
\sY^2&=&\frac{\partial^2}{\partial\dot{\theta}^c{}^2}+
\ctg\dot{\theta}^c\frac{\partial}{\partial\dot{\theta}^c}+
\frac{1}{\sin^2\dot{\theta}^c}\left[
\frac{\partial^2}{\partial\dot{\varphi}^c{}^2}-
2\cos\dot{\theta}^c\frac{\partial}{\partial\dot{\varphi}^c}
\frac{\partial}{\partial\dot{\psi}^c}+
\frac{\partial^2}{\partial\dot{\psi}^c{}^2}\right].\label{KO2}
\end{eqnarray}
Matrix elements of 
unitary irreducible representations\index{representation!unitary irreducible}
of the Lorentz
group are eigenfunctions of the operators (\ref{KO2}):
\begin{eqnarray}
\left[\sX^2+l(l+1)\right]\fM^l_{mn}(\varphi^c,\theta^c,\psi^c)&=&0,\nonumber\\
\left[\sY^2+\dot{l}(\dot{l}+1)\right]\fM^{\dot{l}}_{\dot{m}\dot{n}}
(\dot{\varphi}^c,\dot{\theta}^c,\dot{\psi}^c)&=&0,\label{EQ}
\end{eqnarray}
where
\begin{eqnarray}
\fM^l_{mn}(\varphi^c,\theta^c,\psi^c)&=&e^{-i(m\varphi^c+n\psi^c)}
Z^l_{mn}(\theta^c),\nonumber\\
\fM^{\dot{l}}_{\dot{m}\dot{n}}(\dot{\varphi}^c,\dot{\theta}^c,\dot{\psi}^c)&=&
e^{-i(\dot{m}\dot{\varphi}^c+\dot{n}\dot{\varphi}^c)}
Z^{\dot{l}}_{\dot{m}\dot{n}}(\dot{\theta}^c).\label{HF3'}
\end{eqnarray}
Substituting the functions (\ref{HF3'}) into (\ref{EQ}) and
taking into account the operators (\ref{KO2}), we obtain a complex analog
of the Legendre equations:
\begin{eqnarray}
\left[(1-z^2)\frac{d^2}{dz^2}-2z\frac{d}{dz}-
\frac{m^2+n^2-2mnz}{1-z^2}+l(l+1)\right]Z^l_{mn}&=&0,\label{Leg1}\\
\left[(1-\overset{\ast}{z}{}^2)\frac{d^2}{d\overset{\ast}{z}{}^2}-
2\overset{\ast}{z}\frac{d}{d\overset{\ast}{z}}-
\frac{\dot{m}^2+\dot{n}^2-2\dot{m}\dot{n}\overset{\ast}{z}}
{1-\overset{\ast}{z}{}^2}+\dot{l}(\dot{l}+1)\right]
Z^{\dot{l}}_{\dot{m}\dot{n}}&=&0,\label{Leg2}
\end{eqnarray}
where $z=\cos\theta^c$ and $\overset{\ast}{z}=\cos\dot{\theta}^c$.
The latter equations have three singular points $-1$, $+1$, $\infty$.
Solutions of (\ref{Leg1}) have the form
\begin{multline}
Z^l_{mn}=
\sum^l_{k=-l}i^{m-k}
\sqrt{\Gamma(l-m+1)\Gamma(l+m+1)\Gamma(l-k+1)\Gamma(l+k+1)}\times\\
\cos^{2l}\frac{\theta}{2}\tg^{m-k}\frac{\theta}{2}\times\\[0.2cm]
\sum^{\min(l-m,l+k)}_{j=\max(0,k-m)}
\frac{i^{2j}\tg^{2j}\dfrac{\theta}{2}}
{\Gamma(j+1)\Gamma(l-m-j+1)\Gamma(l+k-j+1)\Gamma(m-k+j+1)}\times\\[0.2cm]
\sqrt{\Gamma(l-n+1)\Gamma(l+n+1)\Gamma(l-k+1)\Gamma(l+k+1)}
\ch^{2l}\frac{\tau}{2}\tnh^{n-k}\frac{\tau}{2}\times\\[0.2cm]
\sum^{\min(l-n,l+k)}_{s=\max(0,k-n)}
\frac{\tnh^{2s}\dfrac{\tau}{2}}
{\Gamma(s+1)\Gamma(l-n-s+1)\Gamma(l+k-s+1)\Gamma(n-k+s+1)}.\label{HS}
\end{multline}
We will call the functions $Z^l_{mn}$ in (\ref{HS}) as
{\it hyperspherical functions}\footnote{The hyperspherical functions (or hyperspherical
harmonics) are known in mathematics for a long time 
(see, for example, Ref. 65). These functions are generalizations
of the three-dimensional spherical functions on the case of $n$-dimensional
euclidean spaces. For that reason we retain this name (hyperspherical
functions) for the case of pseudo-euclidean spaces.}. 
The functions $Z^l_{mn}$ can be written via the 
hypergeometric series as follows:
\begin{multline}
Z^l_{mn}=\cos^{2l}\frac{\theta}{2}\ch^{2l}\frac{\tau}{2}
\sum^l_{k=-l}i^{m-k}\tg^{m-k}\frac{\theta}{2}
\tnh^{n-k}\frac{\tau}{2}\times\\[0.2cm]
\hypergeom{2}{1}{m-l+1,1-l-k}{m-k+1}{i^2\tg^2\dfrac{\theta}{2}}
\hypergeom{2}{1}{n-l+1,1-l-k}{n-k+1}{\tnh^2\dfrac{\tau}{2}}.\label{HS1}
\end{multline}
Therefore, matrix elements are expressed by means of the function
({\it a generalized 
hyperspherical function})
\begin{equation}\label{HS2}
\fT^l_{mn}(\mathfrak{g})=e^{-m(\epsilon+i\varphi)}Z^l_{mn}(\cos\theta^c)
e^{-n(\varepsilon+i\psi)},
\end{equation}
where
\begin{equation}\label{HS3}
Z^l_{mn}(\cos\theta^c)=
\sum^l_{k=-l}P^l_{mk}(\cos\theta)\mathfrak{P}^l_{kn}(\ch\tau),
\end{equation}
here $P^l_{mn}(\cos\theta)$ is a 
generalized spherical 
function on the
group $SU(2)$ (see Ref. 66), and $\mathfrak{P}^l_{mn}$ is an analog of
the generalized spherical function for the group $QU(2)$ (so-called
Jacobi function \cite{Vil68}). $QU(2)$ is a group of quasiunitary
unimodular matrices of second order. As well as the group $SU(2)$, the
group $QU(2)$ is one of the real forms of $SL(2,\C)$
($QU(2)$ is noncompact).
Other designation of this group is
$SU(1,1)$ known also as three-dimensional Lorentz group (this
group is isomorphic to $SL(2,\R)$).
Associated hyperspherical functions are derived from (\ref{HS2}) at $n=0$.
They have the form
\[
\fM^m_l(\fg)=e^{m(\epsilon+i\varphi)}Z^m_l(\cos\theta^c).
\]


\begin{thebibliography}{0}
\bibitem{Wa29} B. L. van der Waerden, {\it Nachr. d. Ces. d.
Wiss. G\"{o}ttingen}, 100 (1929).
\bibitem{BJ53} W. L. Bade, H. Jehle, {\it Rev. Mod. Phys.} 
{\bf 25}, 714 (1953).
\bibitem{LU31} O. Laporte, G. E. Uhlenbeck, {\it Phys. Rev.} {\bf 37}, 1380 (1931).
\bibitem{Cam90} A. A. Campolattaro, {\it Int. J. Theor. Phys.} {\bf 29}, 141, 477 (1990).
\bibitem{VR93} J. Vaz, Jr., W. A. Rodrigues, Jr., {\it Int. J. Theor.
Phys.} {\bf 32}, 945(1993).
\bibitem{Gsp02} A. Gsponer, {\it Int. J. Theor. Phys.} {\bf 41}, 689 (2002).
\bibitem{Rum36} Yu. B. Rumer, {\it Spinorial Analysis} (Moscow, 1936).
\bibitem{Maj} E. Majorana, {\it Scientific Papers}, unpublished, deposited at
the ``Domus Galileana'', Pisa, quaderno {\bf 2}, p.101/1; {\bf 3}, p.11, 160;
{\bf 15}, p.16;{\bf 17}, p.83, 159.
\bibitem{Opp31} J. R. Oppenheimer, {\it Phys. Rev.} {\bf 38}, 725 (1931).
\bibitem{Gup50} S. N. Gupta, {\it Proc. Phys. Soc.} {\bf A63}, 681 (1950).
\bibitem{Ble50} K. Bleuler, {\it Helv. Phys. Acta} {\bf 23}, 567 (1950).
\bibitem{Arc55} W. J. Archibald, {\it Can. J. Phys.} {\bf 33}, 565 (1955).
\bibitem{Blu57} S. A. Bludman, {\it Phys. Rev.} {\bf 107}, 1163 (1957).
\bibitem{Ohm56} T. Ohmura, {\it Prog. Theor. Phys.} {\bf 16}, 684 (1956).
\bibitem{Goo57} R. H. Good, {\it Phys. Rev.} {\bf 105}, 1914 (1957).
\bibitem{Lom58} J. S. Lomont, {\it Phys. Rev.} {\bf 111}, 1710 (1958).
\bibitem{Bor58} A. A. Borhgardt, {\it Sov. Phys. JETP} {\bf 34}, 334 (1958).
\bibitem{Mos59} H. E. Moses, {\it Phys. Rev.} {\bf 113}, 1670 (1959).
\bibitem{SS62} M. Sachs, S. L. Schwebel, {\it J. Math. Phys.} {\bf 3}, 843 (1962).
\bibitem{MRB74} R. Mignani, E. Recami, M. Baldo, {\it Lettere al Nuovo
Cimento} {\bf 11}, 568 (1974).
\bibitem{Bac76} H. Bacry, {\it Nuov. Cim.} {\bf A32}, 448 (1976).
\bibitem{DaS79} A. Da Silveira, {\it Z. Naturforsch} {\bf A34}, 646 (1979).
\bibitem{Gia85} E. Giannetto, {\it Lettere al Nuovo Cimento} {\bf 44}, 140 (1985).
\bibitem{Ljo88} K. Ljolje, {\it Fortschr. Phys.} {\bf 36}, 9 (1988).
\bibitem{Sal90} H. Sallhofer, {\it Z. Naturforsch} {\bf A45}, 1361 (1990).
\bibitem{Sim91} V. M. Simulik, {\it Theor. Math. Phys.} {\bf 87}, 386 (1991).
\bibitem{Ina94} T. Inagaki, {\it Phys. Rev} {\bf A49}, 2839 (1994).
\bibitem{Bir94} I. Bialynicki-Birula, {\it Acta Phys. Pol.} {\bf A86}, 97 (1994). 
\bibitem{Sip95} J. F. Sipe, {\it Phys. Rev.} {\bf A52}, 1875 (1995).
\bibitem{Bru95} S. Bruce, {\it Nuov. Cim.} {\bf B110}, 115 (1995).
\bibitem{Dvo97} V. V. Dvoeglazov, {\it Nuov. Cim.} {\bf 112}, 847 (1997).
\bibitem{Ger98} A. Gersten, {\it Found. Phys. Lett.} {\bf 12}, 291 (1998).
\bibitem{Esp98} S. Esposito, {\it Found. Phys.} {\bf 28}, 231 (1998).
\bibitem{Fin55} D. Finkelstein, {\it Phys. Rev.} {\bf 100}, 924 (1955).
\bibitem{BK69} H. Bacry, A. Kihlberg, {\it J. Math. Phys.} {\bf 10}, 2132 (1969).
\bibitem{Lur64} F. Lur\c{c}at, {\it Physics} {\bf 1}, 95 (1964).
\bibitem{Kih70} A. Kihlberg, {\it Ann. Inst. Henri Poincar\'{e}} {\bf 13}, 
57 (1970).
\bibitem{BF74} C. P. Boyer, G. N. Fleming, {\it J. Math. Phys.}
{\bf 15}, 1007 (1974).
\bibitem{Aro76} H. Arod\'{z}, {\it Acta Phys. Pol.} {\bf B7},
177 (1976).
\bibitem{Tol96} M. Toller, {\it J. Math. Phys.} {\bf 37}, 2694 (1996).
\bibitem{KLS95} S. M. Kuzenko, S. L. Lyakhovich, A. Yu. Segal, 
{\it Int. J. Mod. Phys.} {\bf A10}, 1529 (1995).
\bibitem{LSS96} S. L. Lyakhovich, A. Yu. Segal, A. A. Sharapov,
{\it Phys. Rev.} {\bf D54}, 5223 (1996).
\bibitem{Dre97} W. Drechsler, {\it J. Math. Phys.} {\bf 38}, 5531 (1997).
\bibitem{DG99} A. A. Deriglazov, D. M. Gitman, {\it Mod. Phys. Lett.} 
{\bf A14}, 709 (1999).
\bibitem{GS01} D. M. Gitman, A. L. Shelepin, 
{\it Int. J. Theor. Phys.} {\bf 40}, 603 (2001).
\bibitem{GL01} J. -Y. Grandpeix, F. Lur\c{c}at, {\it Found. Phys.} {\bf32},
109 (2002); ibid. {\bf 32}, 133 (2002).
\bibitem{BBTD88} L. C. Biedenharn, H. W. Braden, P. Truini, H. van Dam,
{\it J. Phys. A: Math. Gen.} {\bf 21},
3593 (1988).
\bibitem{Var034} V. V. Varlamov, {\it Hyperspherical Functions and
Harmonic Analysis on the Lorentz Group}, to appear in
``Progress in Mathematical Physics Research'' (Nova Science Publishers,
New York).
\bibitem{RF} Yu. B. Rumer, A. I. Fet, {\it Group Theory and Quantized Fields}
(Nauka, Moscow, 1977).
\bibitem{Nai58} M. A. Naimark, {\it Linear Representations of the Lorentz Group}
(Pergamon, London, 1964).
\bibitem{Vas96} M. A. Vasiliev, {\it Int. J. Mod. Phys.} {\bf D5}, 763 (1996).
\bibitem{Arn} V. I. Arnold, {\it Mathematical Methods of Classical Mechanics}
(Nauka, Moscow, 1989).
\bibitem{Var022} V. V. Varlamov, {\it Hadronic J.} {\bf 25}, 481 (2002).
\bibitem{AE93} D.V. Ahluwalia, D.J. Ernst, {\it Int. J. Mod. Phys.} {\bf E2}, 397 (1993).
\bibitem{Dvo96} V.V. Dvoeglazov, {\it Nuov. Cim.}
{\bf B111}, 483 (1996).
\bibitem{Ryd85} L. Ryder, {\it Quantum Field Theory} (Cambridge University Press, Cambridge, 1985).
\bibitem{NW49} T. D. Newton, E. P. Wigner, {\it Rev. Mod. Phys.} {\bf 21}, 400 (1949).
\bibitem{Vil68} N.Ya. Vilenkin, {\it Special Functions and the Theory of Group
Representations} (AMS, Providence, 1968).
\bibitem{Pet69} A. Z. Petrov, {\it Einstein Spaces} (Pergamon Press, Oxford,
1969).
\bibitem{HS70} M. Huszar, J. Smorodinsky, {\it Preprint JINR} No. E2-5020, Dubna (1970).
\bibitem{SH70} Ya.A. Smorodinsky, M. Huszar, {\it Teor. Mat. Fiz.}
{\bf 4}, 328 (1970).
\bibitem{Var031} V. V. Varlamov, {\it Int. J. Theor. Phys.} {\bf 42}, 
583 (2003).
\bibitem{Pen84} R. Penrose, W. Rindler, {\it Spinors and Space-Time} 
(Cambridge University Press, Cambridge, 1984).
\bibitem{Var035} V. V. Varlamov, {\it Relativistic wavefunctions on the
Poincar\'{e} group}, preprint math-ph/0308038.
\bibitem{Bat2} H. Bateman, A. Erd\'{e}lyi, {\it Higher Transcendental Functions, 
vol. II}
(Mc Grow-Hill Book Company, New York, 1953).
\bibitem{GMS} I. M. Gel'fand, R. A. Minlos, Z. Ya. Shapiro, {\it Representations
of the Rotation and Lorentz Groups and their Applications} (Pergamon Press,
Oxford, 1963).
\end{thebibliography}
\end{document}